\documentclass[a4paper,10pt,twoside]{cpc-hepnp}

\usepackage{multicol}
\usepackage{graphicx}
\usepackage{booktabs}
\usepackage{amssymb,bm,mathrsfs,bbm,amscd}
\usepackage[tbtags]{amsmath}
\usepackage{lastpage}

\begin{document}
%\begin{CJK*}{GBK}{song}

\fancyhead[co]{\footnotesize ZHANG Wei et al:
Octupole deformation for Ba isotopes in a \\
reflection-asymmetric relativistic mean-field approach}

%\footnotetext[0]{Received xx February 2010}

\title{Octupole deformation for Ba isotopes in a
reflection-asymmetric relativistic mean-field approach
\thanks{Supported by
Foundation of He'nan Educational Committee (200614003),
Young Backbone Teacher Support Program of He'nan Polytechnic University,
China Postdoctoral Science Foundation,
Major State Basic Research Developing Program (2007CB815000), and
National Natural Science Foundation of China (10775004, 10975007, 10975008)}
}

\author{%
      ZHANG Wei(ÕÅì¿)$^{1,2}$%
\quad LI Zhi-Pan(ÀîÖ¾ÅÊ)$^{2}$%
\quad ZHANG Shuang-Quan(ÕÅË«È«)$^{2;1)}$\email{sqzhang@pku.edu.cn}%
}
\maketitle

\address{%
$^1$ School of Electrical Engineering and Automation,
He'nan Polytechnic University, Jiaozuo 454003, China \\
$^2$ State Key Laboratory of Nuclear Physics and Technology, School of Physics,
Peking University, Beijing 100871, China\\
}

\begin{abstract}
The potential energy surfaces of even-even $^{142-156}$Ba are
investigated in the constrained reflection-asymmetric relativistic
mean-field approach with parameter set PK1. It is shown that for the ground states,
$^{142}$Ba is near spherical,$^{156}$Ba well quadrupole-deformed, and
in between $^{144-154}$Ba octupole deformed.
In particular, the nuclei $^{148,150}$Ba with $N=92,94$ have the largest octupole deformations.
By including the octupole degree of freedom, energy gaps $N=88$, $N=94$ and $Z=56$
near Fermi surfaces for the single-particle levels in $^{148}$Ba
with $\beta_2\sim 0.26$ and $\beta_3\sim 0.17$ are found.
Furthermore, the performance of the octupole deformation driving pairs
($\nu 2f_{7/2}$, $\nu 1i_{13/2}$) and ($\pi 2d_{5/2}$, $\pi 1h_{11/2}$) is demonstrated
by analyzing the single-particle levels near Fermi surfaces in $^{148}$Ba.
\end{abstract}

\begin{keyword}
reflection-asymmetric, relativistic mean-field,
octupole deformation, single-particle levels
\end{keyword}

\begin{pacs}
21.10.-k, 21.60.Jz, 27.60.+j, 27.70.+q
% 21.10.-k Properties of nuclei; nuclear energy levels properties
% 21.10.Dr Binding energies and masses
% 21.10.Pc Single-particle levels and strength functions
% 21.10.Gv Nuclear deformation ¡ªnucleon distribution
% 21.60.Jz Nuclear Density Functional Theory and extensions (includes Hartree¨CFock and random-phase approximations)
% 27.60.+j 90<=A<=149
% 27.70.+q 150<=A<=189
\end{pacs}

%\begin{multicols}{2}
%\Large

\section{Introduction}

In the last decades, the phenomena related to octupole deformation have
received wide attention.
Normally the regions of nuclei with strong octupole correlations
correspond to either the proton or neutron numbers close to 34
($1g_{9/2} \leftrightarrow 2p_{3/2}$ coupling), 56 ($1h_{11/2}
\leftrightarrow 2d_{5/2}$ coupling), 88 ($1i_{13/2} \leftrightarrow
2f_{7/2}$ coupling), and 134 ($1j_{15/2} \leftrightarrow 2g_{9/2}$
coupling)~\cite{Butler1996}.

Extensive efforts have been made for understanding the structure of
neutron-rich nuclei around $Z\sim56$ and $N\sim88$.
Experimentally in this region, many octupole deformed bands have been
identified and extended to higher spin,
such as in $^{139}$Xe~\cite{Zhu1997-1,Luo2002},
in even-even $^{140-148}$Ba~\cite{Phillips1986,Zhu1995,Zhu1997-2,Urban1997},
in $^{144,146}$Ce~\cite{Zhu1998}, and in $^{145,147}$La~\cite{Zhu1999}.
On the theoretical side, a variety of approaches have been
applied to investigate the role of octupole degree of freedom in this nuclear region.
The Woods-Saxon-Bogoliubov cranking model is used to study the shapes of
rotating Xe, Ba, Ce, Nd, and Sm nuclei with $N=84-94$ and the expectations of
octupole-deformed mean-fields at low and medium spins are confirmed~\cite{Nazarewicz1992}.
Based on the Hartree-Fock plus BCS approximation and
adiabatic time-dependent Hartree-Fock plus zero point energy in the cranking approximation,
the energy splitting and $B(E1)$ transition are well described for
$^{140}$Ba, $^{142-150}$Ce, $^{144-152}$Nd, and $^{146-154}$Sm~\cite{Egido1992}.
In the {\it spdf} interacting boson model,
good agreement of the calculated low-lying energy spectra and transition rates
with data is obtained for $^{144}$Ba and $^{146}$Ba~\cite{Liu1994}.
The reflection-asymmetric shell model is applied to describe the octupole deformed bands in neutron-rich $^{142}$Ba and $^{145}$Ba, and good agreement with the experimental data is obtained~\cite{Chen2005}.

During the past years, the Relativistic Mean-Field (RMF) theory
~\cite{Ring1996,Vretenar2005,Meng2006}
has achieved great success in describing many nuclear phenomena
related to stable nuclei~\cite{Ring1996}, exotic nuclei~\cite{Meng1996,Meng1998}
as well as supernova and neutron stars~\cite{Glendenning2000}.
Recently, the Reflection-ASymmetric Relativistic Mean-Field
(RAS-RMF) approach considering the octupole degree of freedom is developed and applied
to the well-known octupole deformed nucleus $^{226}$Ra~\cite{Geng2007}, and La isotopes~\cite{Wang2009}.
In Ref.~\cite{Zhang2009}, the RAS-RMF approach has been applied to investigate the Potential Energy Surfaces (PESs) of even-even $^{146-156}$Sm isotopes in the ($\beta_2$,~$\beta_3$) plane,
and it is suggested that the critical-point candidate nucleus $^{152}$Sm marks the shape/phase
transition not only from $U(5)$ to $SU(3)$ symmetry, but also from
the octupole deformed ground state in $^{150}$Sm  to quadrupole
deformed ground state in $^{154}$Sm.
%Both the binding energy and deformation are well reproduced.
Therefore, it is interesting to investigate the Ba isotopes
in the RAS-RMF approach.

In this paper, the RAS-RMF approach will be applied to investigate
the potential energy surfaces of even-even $^{142-156}$Ba
isotopes in the ($\beta_2$,~$\beta_3$) plane, and
the single-particle levels near Fermi surfaces
for the nucleus $^{148}$Ba will be studied.

\section{Formalism}

The basic ansatz of the RMF theory is a Lagrangian density where
nucleons are described as Dirac particles which interact via the
exchange of various mesons and the photon. The mesons considered are
the isoscalar-scalar $\sigma$, the isoscalar-vector $\omega$ and the
isovector-vector $\rho$. The effective Lagrangian density
reads~\cite{Serot1986},

\begin{eqnarray} \nonumber
 \cal L&=&\bar\psi \bigg[
i\gamma^\mu\partial_\mu-M-g_\sigma\sigma-g_\omega\gamma^\mu\omega_\mu\\
\nonumber
&&-g_\rho\gamma^\mu \vec\tau \cdot \vec\rho_\mu - e\gamma^\mu\dfrac{1-\tau_3}{2}A_\mu \bigg] \psi\\
\nonumber
&&+\dfrac{1}{2}\partial^\mu\sigma\partial_\mu\sigma-\dfrac{1}{2} m_\sigma^2\sigma^2
-\dfrac{1}{3}g_2\sigma^3-\dfrac{1}{4}g_3\sigma^4\\
\nonumber
&&-\dfrac{1}{4}\Omega^{\mu\nu}\Omega_{\mu\nu}+\dfrac{1}{2}m_\omega^2\omega^\mu\omega_\mu
+\dfrac{1}{4}c_3 (\omega^\mu\omega_\mu)^2\\
\nonumber
&&-\dfrac{1}{4}\vec R^{\mu\nu}\cdot\vec R_{\mu\nu}+\dfrac{1}{2}m_\rho^2\vec\rho^\mu\cdot\vec\rho_\mu\\
&&-\dfrac{1}{4}F^{\mu\nu}F_{\mu\nu},
 \label{Lagrangian}
\end{eqnarray}
in which the field tensors for the vector mesons and the photon are
respectively defined as,
\begin{eqnarray}
\left\{
\begin{array}{lll}
   \Omega_{\mu\nu}   &=& \partial_\mu\omega_\nu-\partial_\nu\omega_\mu, \\
   {\vec R}_{\mu\nu} &=& \partial_\mu{\vec \rho}_\nu
                        -\partial_\nu{\vec \rho}_\mu, \\
   F_{\mu\nu}        &=& \partial_\mu A_\nu-\partial_\nu  A_\mu.
\end{array}   \right.
\end{eqnarray}

Using the classical variational principle, one can obtain
the Dirac equation for the nucleons and the Klein-Gordon
equations for the mesons. To solve these
equations, we employ the basis expansion method,
which has been widely used in both the non-relativistic
and relativistic mean-field models. For axial-symmetric
reflection-asymmetric systems,
the spinors are expanded in terms of the eigenfunctions of the
Two-Center Harmonic-Oscillator (TCHO) potential
\begin{eqnarray}
V(r_\perp,z)= \dfrac{1}{2} M \omega_\perp^2 r_\perp^2 +\left\{
\begin{array}{ll}
   \dfrac{1}{2} M \omega_1^2(z+z_1)^2, & z<0\\
   &\\
   \dfrac{1}{2} M \omega_2^2(z-z_2)^2, & z\geqslant0\\
\end{array}   \right.
\end{eqnarray}
where $M$ is the nucleon mass, $z_1$ and $z_2$ (real, positive)
represent the distances between the centers of
the spheroids and their intersection plane, and $\omega_1$($\omega_2$)
are the corresponding oscillator frequencies for $z < 0$ ($z \geqslant 0$)~\cite{Geng2007}.
The TCHO basis has been widely used in the studies of fission, fusion,
heavy-ion emission, and various cluster phenomena~\cite{Greiner1994}.
By setting proper asymmetric parameters,
the major and the $z$-axis quantum numbers are real numbers very close to integers,
and the integers are used in the Nilsson-like notation $\Omega$[$N$$n_z$$m_l$] for convenience.
More details can be found in Ref.~\cite{Geng2007}.

The binding energy at a certain deformation is obtained by
constraining the mass quadrupole moment $\langle \hat{Q_2}\rangle $
to a given value $\mu_2$~\cite{Ring1980}, i.e.,
 \begin{equation}
  \langle H'\rangle ~=~\langle H\rangle  +  \dfrac{1}{2}C(\langle \hat{Q_2}\rangle -\mu_2)^2
 \end{equation}
where $C$ is the curvature constant parameter, and $\mu_2$ is
the given quadrupole moment. The expectation value of $\hat{Q_2}$ is
 $\langle \hat{Q_2}\rangle =\langle \hat{Q_2}\rangle _n+\langle \hat{Q_2}\rangle
 _p$ with
 $\langle\hat{Q_2}\rangle _{n,p}= \langle 2 r^2 P_2(\cos\theta)\rangle _{n,p}$.
The deformation parameter $\beta_2$ is related to $\langle
\hat{Q_2}\rangle $ by, $\langle \hat{Q_2}\rangle  =
\dfrac{3}{\sqrt{5\pi}} Ar^2\beta_2$ with $r = R_0 A^{1/3}$
($R_0=1.2$ fm) and $A$ the mass number. The octupole moment
constraint can also be applied similarly with
$\langle\hat{Q_3}\rangle =\langle \hat{Q_3}\rangle _n+\langle
\hat{Q_3}\rangle _p$, $\langle\hat{Q_3}\rangle _{n,p}= \langle 2 r^3
P_3(\cos\theta)\rangle _{n,p}$, and $\langle\hat{Q_3}\rangle  =
\dfrac{3}{\sqrt{7\pi}} Ar^3\beta_3$. By constraining the quadruple
moment and octupole moment simultaneously, the total energy surface
in ($\beta_2$,~$\beta_3$) plane can be obtained.

\section{Results and discussion}

The properties of even-even $^{142-156}$Ba are calculated
in the constrained RAS-RMF approach with parameter set PK1~\cite{Long2004}.
The TCHO basis with 16 major shells for both fermions and bosons is used.
The pairing correlation is treated by the BCS approximation with
a constant pairing gap $\Delta=11.2/\sqrt{A}$ MeV.

To investigate the shape evolution in $^{142-156}$Ba,
the total energies as functions of $\beta_2$ and $\beta_3$ have been analyzed.
As an example, Fig.~\ref{fig:pes} displays the contour plots for $^{142}$Ba and $^{148}$Ba.
It is found that for the ground states,
$^{142}$Ba is near spherical without octupole deformation,
$^{144-154}$Ba octupole deformed and $^{156}$Ba well quadrupole-deformed.
%Because of the equivalence between the states with positive and negative $\beta_3$,
%the contour plots have up-down symmetry in the ($\beta_2$,~$\beta_3$) plane.
In detail, for $^{144-156}$Ba,
the quadrupole deformation $\beta_2$ of the global octupole minimum gradually increases
with increasing neutron numbers.
On the other hand,
the octupole deformation $\beta_3$ of the global octupole minimum gradually increases
for $^{144,146,148}$Ba, and decreases for $^{150,152,154}$Ba,
and until $^{156}$Ba the global minimum is well deformed with little octupole deformation.
A soft area covering the global octupole minimum and the saddle point
at $\beta_3$=0 appears in $^{144}$Ba, and develops in $^{146-154}$Ba.
Furthermore, the energy difference between the global octupole minimum and
the saddle point increases from $^{144}$Ba to $^{148}$Ba,
while it decreases from $^{150}$Ba to $^{156}$Ba.

%\end{multicols}
%\ruleup
\begin{center}
\includegraphics[scale=0.4]{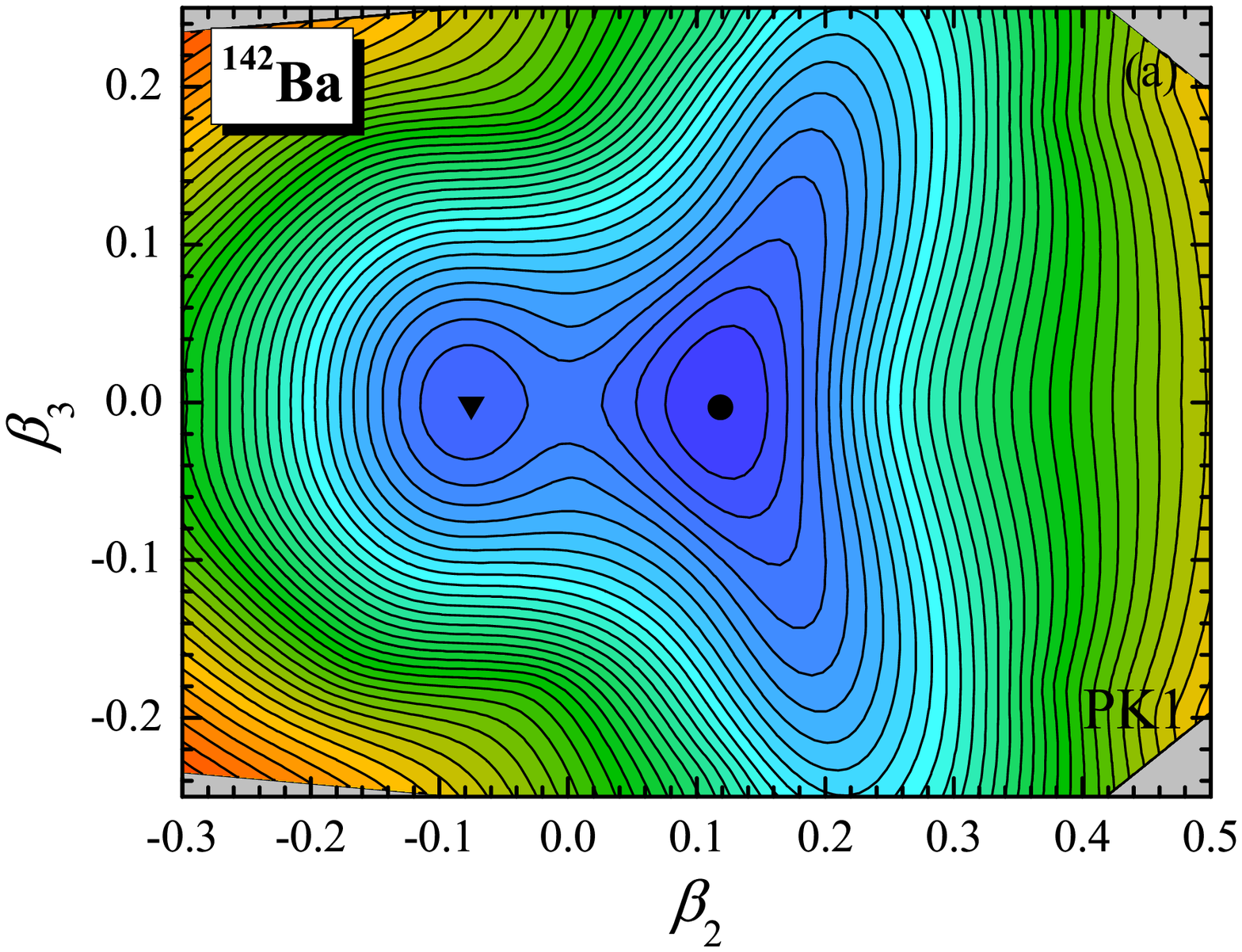}
\includegraphics[scale=0.4]{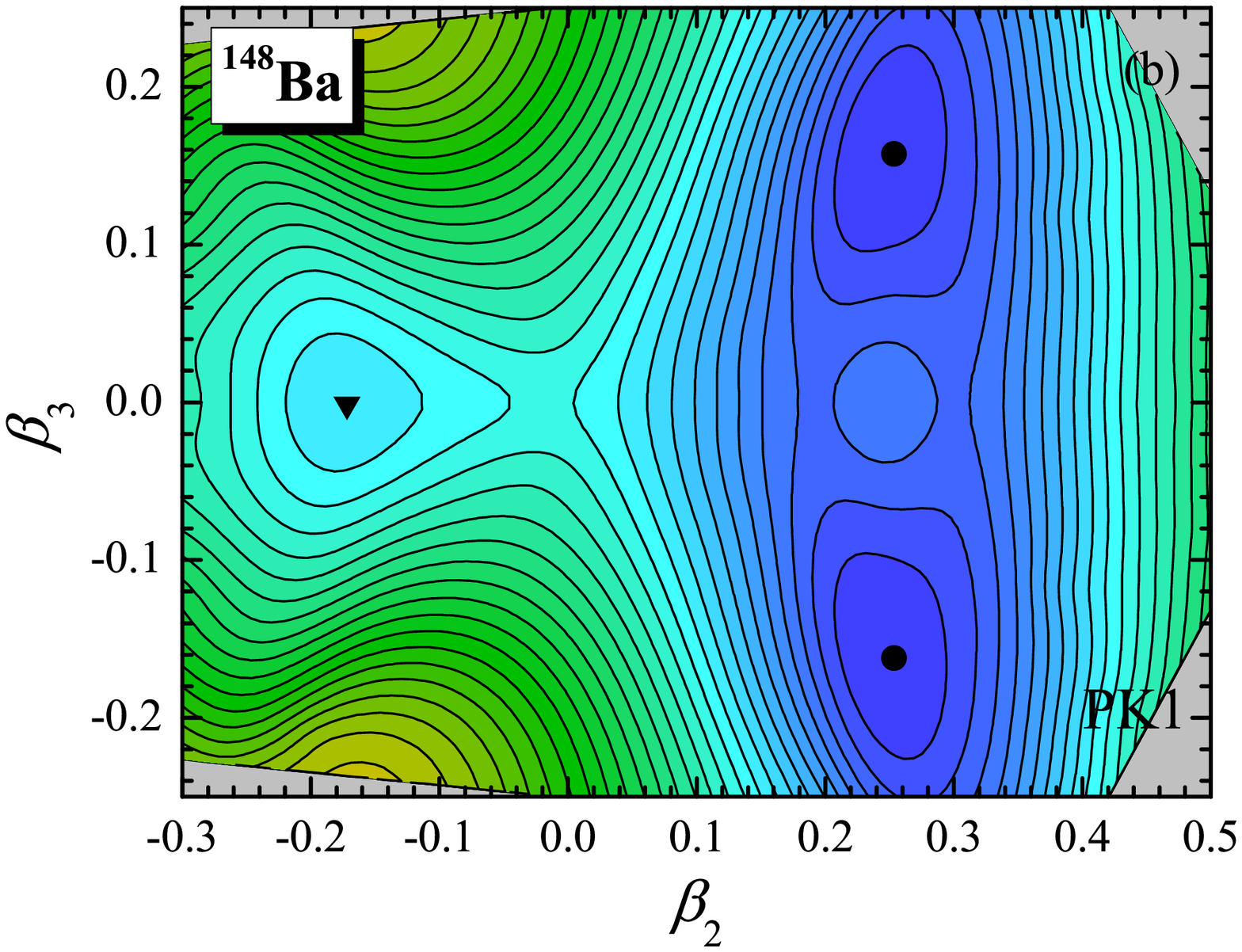}
\figcaption{\label{fig:pes}
(Color online)
The contour plots of total energies for even-even $^{142}$Ba and $^{148}$Ba in ($\beta_2$,~$\beta_3$) plane
obtained in RAS-RMF approach with PK1 and constant-$\Delta$ pairing.
The energy separation between contour lines is 0.5 MeV.
The global minimum and other local minima are denoted by ``$\bullet$" and ``$\blacktriangledown$" respectively.
}
\end{center}
%\ruledown
%\begin{multicols}{2}

The binding energy, quadrupole and octupole deformations
are summarized for the ground states of even-even $^{142-156}$Ba in Table~\ref{tab:be}.
The binding energies are well reproduced within 0.3\%. Moreover,
excellent agreement is obtained for the quadrupole deformations.
It's indicated that the ground states of even-even $^{144-154}$Ba with $N=88-98$ are octupole deformed
while in the middle the nuclei $^{148,150}$Ba with $N=92-94$ are the most octupole deformed.
It is noted that the suggested octupole Ba nuclei ($N=88-98$) are more neutron-rich than those ($N=86-91$)
predicted in the finite-range droplet model~\cite{Moller1995}.
This conclusion consists with the previous RAS-RMF calculation for La isotopes~\cite{Wang2009}.
%where the existence of the octupole deformation are suggested for
%$^{145-155}$La with $N=88-98$

\begin{center}
\tabcaption{ \label{tab:be}
The total binding energy (in MeV) as well as the quadrupole deformation $\beta_2$
and octupole deformation $\beta_3$ of the ground states of even-even $^{142-156}$Ba obtained in
the constrained RAS-RMF approach with PK1,
in comparison with the available experimental data.}
\footnotesize
\begin{tabular*}{80mm}{@{\extracolsep{\fill}}cccccc}
\toprule
Nucleus & $E^{\rm cal}$ &$\beta_2^{\rm cal}$ &$\beta_3^{\rm cal}$&
$E^{\rm exp}$~\cite{Audi2003} &$\beta_2^{\rm exp}$~\cite{Raman2001} \\
\hline
$^{142}$Ba  &1181.25  &0.12 &0.00 &1180.14  &0.16\\
$^{144}$Ba  &1190.19  &0.20 &0.12 &1190.23  &0.19\\
$^{146}$Ba  &1199.06  &0.23 &0.13 &1199.60  &0.22\\
$^{148}$Ba  &1207.26  &0.26 &0.17 &1208.76  &$-$ \\
$^{150}$Ba  &1215.12  &0.28 &0.17 &1217.55  &$-$ \\
$^{152}$Ba  &1221.86  &0.30 &0.12 &1225.58  &$-$ \\
$^{154}$Ba  &1228.30  &0.32 &0.08 &$-$      &$-$ \\
$^{156}$Ba  &1234.58  &0.33 &0.03 &$-$      &$-$ \\
\bottomrule
\end{tabular*}
\end{center}

To understand the evolution of the octupole deformation,
the microscopic single-particle levels are analyzed.
Fig.~\ref{fig:sp1} shows
the neutron single-particle levels for the states minimized with respect to $\beta_3$ and the states
with $\beta_3$=0 for $\beta_2=0.14 \sim 0.38$ in $^{148}$Ba side by side.
The levels near the Fermi surface are labeled by
Nilsson-like notations $\Omega[Nn_zm_l]$ of the largest component.
In the left panel of Fig.~\ref{fig:sp1}, an energy gap $N=88$
fenced with several levels near the Fermi surface
for $\beta_2=0.20 \sim 0.30$ is found approaching the octupole minimum.
Additionally, a gap at $N=94$ is found for $\beta_2 \sim 0.28$.
For the states with $\beta_3$=0, the neutron gap at $N=88$ is relatively small, and the
gap at $N=94$ does not appear.
Furthermore, the proton single-particle levels for the corresponding states
are shown in Fig.~\ref{fig:sp2}.
An energy gap $Z=56$ appears for the states with octupole deformation approaching the ground state
in the left panel of Fig.~\ref{fig:sp2},
while the energy gap $Z=56$ can not be found near the Fermi surfaces for the states with $\beta_3$=0
in the right panel of Fig.~\ref{fig:sp2}.
Thus the energy gaps $N=88$, $N=94$, and $Z=56$ near Fermi surfaces
are responsible for the octupole minimum of Ba isotopes.
In particular, the energy gaps $N=94$ and $Z=56$ around $\beta_2\sim 0.28 $ are consistent with
the large octupole deformation predicted for $^{148,150}$Ba.

%\end{multicols}
%\ruleup
\begin{center}
%\includegraphics[scale=0.58]{Ba.figures/Ba148.PK1.N.eps}
%\hspace{0.5cm}
%\includegraphics[scale=0.58]{Ba.figures/Ba148.PK1.N.nobeta3.eps}
\includegraphics[scale=0.58]{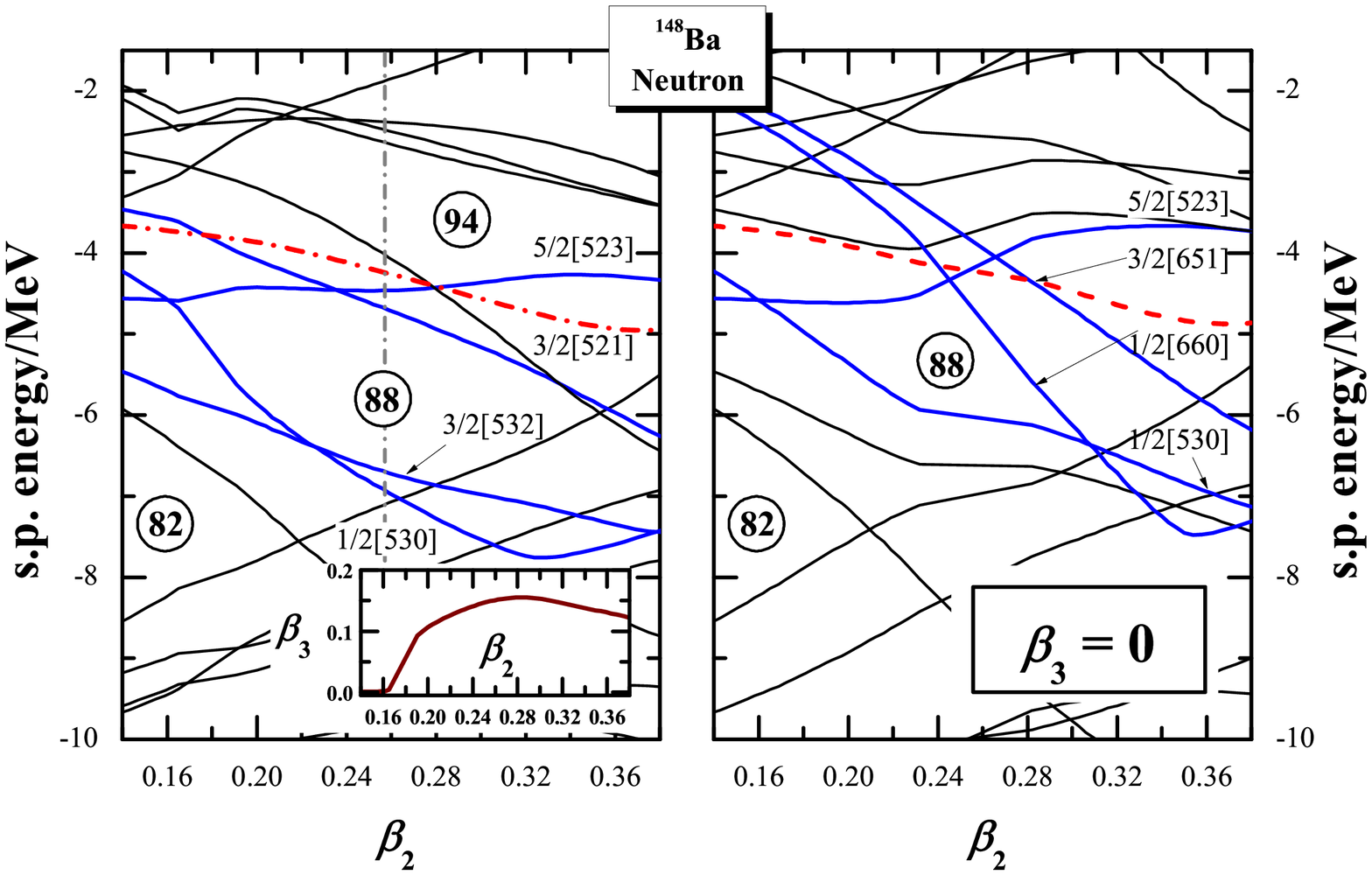}
\figcaption{\label{fig:sp1}
(Color online)
Neutron single-particle levels of $^{148}$Ba in RAS-RMF approach
with PK1 as functions of $\beta_2$ for
states minimized with respect to $\beta_3$ (left panel) and states with $\beta_3 =0$ (right panel).
The dash-dot lines denote the corresponding Fermi surfaces.
The levels near the Fermi surface are labeled by
Nilsson-like notations $\Omega$[$N$$n_z$$m_l$] of the first component.
The corresponding $\beta_3$ are shown in the inset.
The quadrupole deformation of the ground state is indicated by the vertical gray line.
}
\end{center}

\begin{center}
%\includegraphics[scale=0.58]{Ba.figures/Ba148.PK1.P.eps}
%\hspace{0.5cm}
%\includegraphics[scale=0.58]{Ba.figures/Ba148.PK1.P.nobeta3.eps}
\includegraphics[scale=0.58]{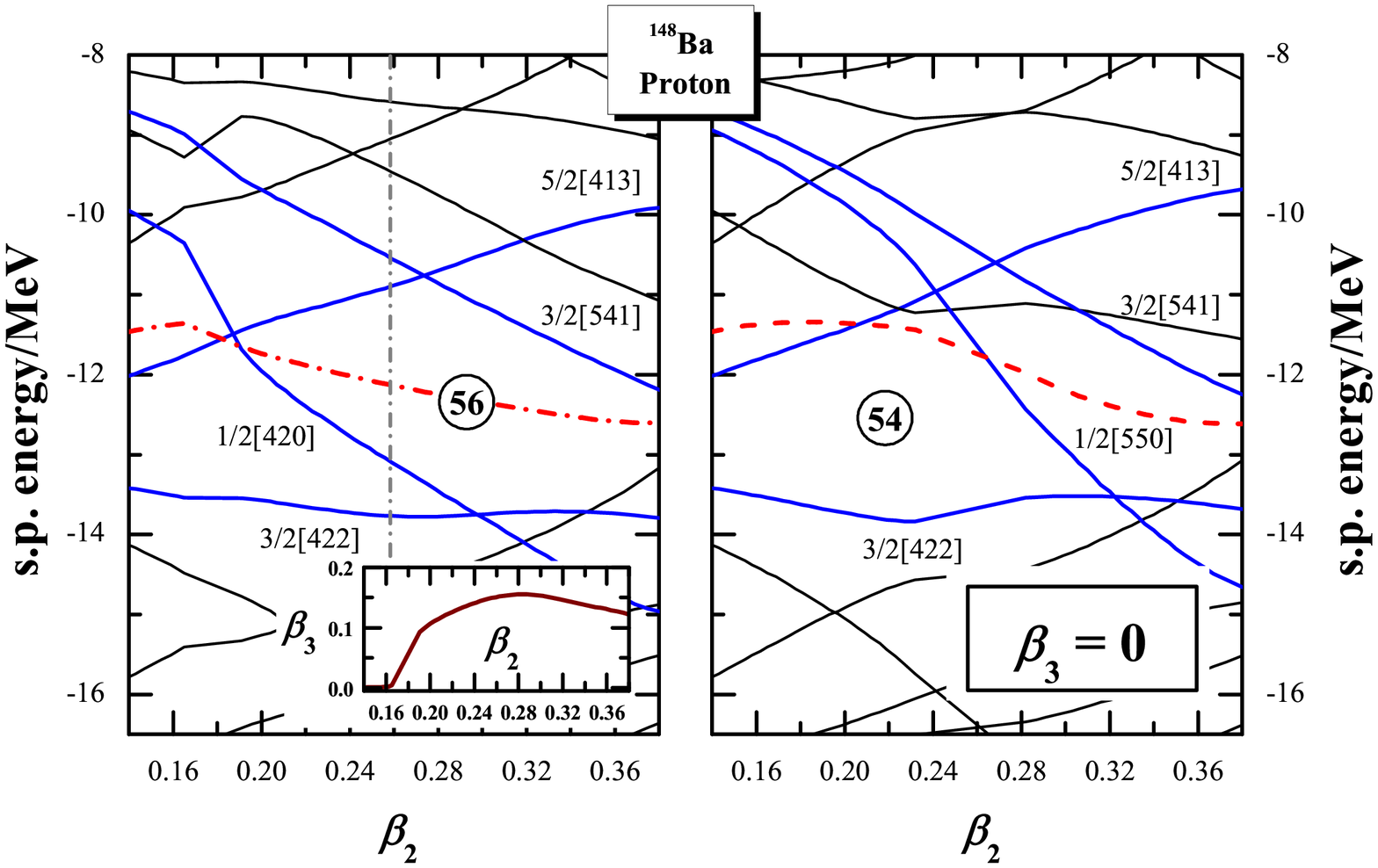}
\figcaption{\label{fig:sp2}
(Color online)
Same as Fig.~\ref{fig:sp1}, but for proton.}
\end{center}

%\ruledown
%\begin{multicols}{2}

It is well-known that for nuclei with $N\sim88$ or $Z\sim56$ the
octupole deformation driving pairs of orbitals include ($\nu
2f_{7/2}$, $\nu 1i_{13/2}$) and ($\pi 2d_{5/2}$, $\pi 1h_{11/2}$),
which in the axially deformed case will be subgrouped as
($\nu1/2[541]$, $\nu1/2[660]$), ($\nu3/2[532]$, $\nu3/2[651]$),
($\nu5/2[523]$,$\nu5/2[642]$), ($\nu7/2[514]$, $\nu7/2[633]$), and
($\pi1/2[431]$, $\pi1/2[550]$), ($\pi3/2[422]$, $\pi3/2[541]$),
($\pi5/2[413]$, $\pi5/2[532]$), respectively. It is interesting to
investigate the performance of such pairs in the single-particle
levels near Fermi surfaces in Figs.~\ref{fig:sp1} and \ref{fig:sp2}.
These levels together with their BCS occupation probabilities and
their ten leading components for the ground state are shown in Tables~\ref{tab:com1}
and ~\ref{tab:com2} respectively.
%3/2[521]&  36.0\%
%3/2[532]&  18.7\%
%3/2[651]&  16.5\%
%3/2[631]&   6.8\%
Taking the level $\nu 3/2[521]$ as an example,
its second (18.7\%) and third (16.5\%) components compose
an octupole deformation driving pair ($\nu3/2[532]$, $\nu3/2[651]$).
Similarly, one can find
the pair ($1/2[541]$, $1/2[660]$) for $\nu 1/2[530]$,
the pair ($3/2[532]$, $3/2[651]$) for $\nu 3/2[532]$,
and the pair ($5/2[523]$, $5/2[642]$) for $\nu 5/2[523]$.
For the proton side,
octupole deformation driving pairs are also found among the components of
single-particle levels near Fermi surfaces at the ground state of $^{148}$Ba.
Therefore both the neutron and the proton driving pairs
play important roles for the octupole minimum in $^{148}$Ba.

%\end{multicols}

%\ruleup
\begin{center}
\tabcaption{ \label{tab:com1}
Single-particle levels near Fermi surface for the ground
state in $^{148}$Ba together with their BCS occupation probabilities
and corresponding contributions from the ten leading components.
The components originating from the octupole deformation driving pairs of orbitals
($\nu 2f_{7/2}$, $\nu 1i_{13/2}$) and ($\pi 2d_{5/2}$, $\pi 1h_{11/2}$) are in bold.
}
%\vspace{-3mm}
\footnotesize
\begin{tabular*}{170mm}{@{\extracolsep{\fill}}c|cc|cc|cc|cc}
\toprule
level &    \multicolumn{2}{c|}{$\nu 1/2[530]$}&                \multicolumn{2}{c|}{$\nu 3/2[532]$}&             \multicolumn{2}{c|}{$\nu 3/2[521]$}&         \multicolumn{2}{c}{$\nu 5/2[523]$}      \\
occu. &    \multicolumn{2}{c|}{0.973}&                         \multicolumn{2}{c|}{0.968}&                      \multicolumn{2}{c|}{0.721}&                  \multicolumn{2}{c}{0.609}\\
1st  comp. & 1/2[530]& 41.0\% &             {\bf 3/2[532]}&{\bf  52.6\%} &          3/2[521]& 36.0\% &      {\bf 5/2[523]}&{\bf  56.4\%} \\
2nd  comp. &{\bf  1/2[541]}&{\bf  15.6\%} &             3/2[541]& 17.6\% &          {\bf 3/2[532]}&{\bf  18.7\%} &      5/2[532]& 12.7\% \\
3rd  comp. &{\bf  1/2[660]}&{\bf   6.2\%} &             3/2[512]&  6.2\% &          {\bf 3/2[651]}&{\bf  16.5\%} &      {\bf 5/2[642]}&{\bf  10.1\%} \\
4th  comp. & 1/2[651]&  5.8\% &             {\bf 3/2[651]}&{\bf   6.0\%} &          3/2[631]&  6.8\% &      5/2[633]&  5.0\% \\
5th  comp. & 1/2[510]&  5.7\% &     3/2[631]&  3.7\% &   3/2[642]&  6.7\% &    5/2[622]&  3.3\% \\
6th  comp. & 1/2[631]&  3.5\% &     3/2[521]&  2.2\% &   3/2[761]&  2.5\% &    5/2[503]&  3.2\% \\
7th  comp. & 1/2[640]&  3.2\% &     3/2[761]&  1.7\% &   3/2[501]&  2.1\% &    5/2[512]&  2.0\% \\
8th  comp. & 1/2[770]&  3.1\% &     3/2[402]&  1.6\% &   3/2[622]&  2.0\% &    5/2[413]&  1.3\% \\
9th  comp. & 1/2[400]&  2.6\% &     3/2[642]&  1.1\% &   3/2[721]&  0.9\% &    5/2[752]&  0.8\% \\
10th comp. & 1/2[521]&  1.9\% &     3/2[321]&  0.9\% &   3/2[752]&  0.8\% &    5/2[613]&  0.5\% \\

\bottomrule
\end{tabular*}%
\end{center}

\begin{center}
\tabcaption{ \label{tab:com2}
Same as Table \ref{tab:com1}, but for proton.}
%\vspace{-3mm}
\footnotesize
\begin{tabular*}{170mm}{@{\extracolsep{\fill}}c|cc|cc|cc|cc}
\toprule
level &    \multicolumn{2}{c|}{$\pi 3/2[422]$}&       \multicolumn{2}{c|}{$\pi 1/2[420]$}&      \multicolumn{2}{c|}{$\pi 5/2[413]$}&      \multicolumn{2}{c}{$\pi 3/2[541]$}   \\
occu. &    \multicolumn{2}{c|}{ 0.936}&               \multicolumn{2}{c|}{ 0.863}&              \multicolumn{2}{c|}{ 0.098}&              \multicolumn{2}{c}{ 0.068}           \\
1st  comp. &{\bf 3/2[422]}&{\bf  69.4\%} &    1/2[420]& 38.6\% &   {\bf 5/2[413]}&{\bf  78.7\%} &  {\bf  3/2[541]}&{\bf  40.8\%} \\
2nd  comp. &3/2[431]& 10.7\% &   {\bf  1/2[550]}&{\bf  14.4\%} &   5/2[422]&  8.9\% &   3/2[521]& 15.3\% \\
3rd  comp. &3/2[402]&  5.1\% &    1/2[530]& 11.0\% &   5/2[523]&  4.2\% &   3/2[411]& 13.9\% \\
4th  comp. &{\bf 3/2[541]}&{\bf   3.8\%} &    1/2[541]&  8.6\% &   {\bf 5/2[532]}&{\bf  2.0\%} &   3/2[532]& 10.2\% \\
5th  comp. &3/2[532]&  2.0\% &    1/2[400]&  4.2\% &   5/2[303]&  1.7\% &  {\bf  3/2[422]}&{\bf   7.1\%} \\
6th  comp. & 3/2[521]&  1.7\% &             1/2[521]&  3.7\% & 5/2[512]&  0.5\% & 3/2[651]&  2.2\% \\
7th  comp. & 3/2[312]&  1.4\% &             1/2[660]&  3.3\% & 5/2[813]&  0.1\% & 3/2[512]&  1.9\% \\
8th  comp. & 3/2[211]&  0.7\% &{\bf  1/2[431]}&{\bf   2.7\%} & 5/2[633]&  0.1\% & 3/2[431]&  1.1\% \\
9th  comp. & 3/2[651]&  0.5\% &             1/2[220]&  1.6\% & 5/2[833]&  0.1\% & 3/2[761]&  0.9\% \\
10th comp. & 3/2[642]&  0.3\% &             1/2[411]&  1.5\% & 5/2[312]&  0.1\% & 3/2[321]&  0.6\% \\

\bottomrule
\end{tabular*}%
\end{center}

\vspace{5mm}
%\ruledown
%\begin{multicols}{2}

\section{Summary}

In conclusion, the PESs of even-even $^{142-156}$Ba in
($\beta_2$,~$\beta_3$) plane are investigated by the constrained RAS-RMF
approach, and the single-particle levels near Fermi surfaces for the
nucleus $^{148}$Ba are studied.
It is shown that for the ground states,
$^{142}$Ba is near spherical without octupole deformation,
$^{144-154}$Ba octupole deformed and $^{156}$Ba well quadrupole-deformed.
%In particular, the nuclei $^{148,150}$Ba with $N=92,94$ have the largest octupole deformations.
The nuclei with largest octupole deformation $\beta_3$ for the ground states
are predicted to be $^{148,150}$Ba ($N=92,94$).

By including the octupole degree of freedom, energy gaps $N=88$, $N=94$, and $Z=56$
near Fermi surfaces are found in single-particle levels approaching the ground state of $^{148}$Ba.
Furthermore, the performance of the octupole deformation driving pairs
($\nu 2f_{7/2}$, $\nu 1i_{13/2}$) and ($\pi 2d_{5/2}$, $\pi 1h_{11/2}$) is demonstrated
by analyzing the components of the single-particle levels near Fermi surfaces.

\acknowledgments{The authors gratefully acknowledge Professor MENG Jie,
Dr. GENG Li-Sheng and LIANG Hao-Zhao for their helpful suggestions and stimulating discussions.}

%\end{multicols}

\vspace{-1mm}
\centerline{\rule{80mm}{0.1pt}}
\vspace{2mm}

%\begin{multicols}{2}

%\end{multicols}

\clearpage

%\end{CJK*}
\end{document}